\documentclass[conference, a4paper]{IEEEtran}
\IEEEoverridecommandlockouts
\usepackage{xspace,amsmath,amssymb,amsfonts,epsfig,subfigure,syntonly}
\usepackage{lipsum}
\usepackage{cite,bm,color,url,textcomp,empheq,boxedminipage}
\usepackage{algorithmicx,algorithm}
\usepackage{indentfirst}
\usepackage{epstopdf,makecell}
\usepackage{empheq}
\usepackage{pifont}
\usepackage{stfloats}
\usepackage[noend]{algpseudocode}

\usepackage{graphicx,graphics}  
\usepackage{multirow,multicol}
\usepackage{psfrag}    
\usepackage{stfloats}
\usepackage{url}
\usepackage{lipsum}

\newtheorem{remark}{\underline{Remark}}[section]

\newcommand{\mv}[1]{\mbox{\boldmath{$ #1 $}}}

\long\def\symbolfootnote[#1]#2{\begingroup
\def\thefootnote{\fnsymbol{footnote}}
\footnote[#1]{#2}\endgroup}
\psfull

\allowdisplaybreaks[4]

\begin{document}
\title{Derivative-Free Placement Optimization for Multi-UAV Wireless Networks with Channel Knowledge Map}

\author{\IEEEauthorblockN{Haoyun Li\IEEEauthorrefmark{1},
Peiming Li\IEEEauthorrefmark{1}\IEEEauthorrefmark{2}\IEEEauthorrefmark{3}, Jie Xu\IEEEauthorrefmark{1}, Junting Chen\IEEEauthorrefmark{1}, and Yong Zeng\IEEEauthorrefmark{3}\IEEEauthorrefmark{4}}
\IEEEauthorblockA{\IEEEauthorrefmark{1}School of Science and Engineering, and Future Network of Intelligence Institute, \\The Chinese University of Hong Kong, Shenzhen, Shenzhen 518172, China\\
\IEEEauthorrefmark{2}School of Information Engineering, Guangdong University of Technology, Guangzhou 510006, China\\
\IEEEauthorrefmark{3}National Mobile Communications Research Laboratory, Southeast University, Nanjing 210096, China\\
\IEEEauthorrefmark{4}Purple Mountain Laboratories, Nanjing 211111, China\\
E-mail: haoyunli@link.cuhk.edu.cn, peiminglee@outlook.com, \{xujie, juntingc\}@cuhk.edu.cn, yong\_zeng@seu.edu.cn}}

\maketitle
\begin{abstract}
    This paper studies a multi-UAV wireless network, in which multiple UAV users share the same spectrum to send individual messages to their respectively associated ground base stations (GBSs). The UAV users aim to optimize their locations to maximize the weighted sum rate. While most existing work considers simplified line-of-sight (LoS) or statistic air-to-ground (A2G) channel models, we exploit the location-specific channel knowledge map (CKM) to enhance the placement performance in practice. However, as the CKMs normally contain discrete site- and location-specific channel data without analytic model functions, the corresponding weighted sum rate function becomes non-differentiable in general. In this case, conventional optimization techniques relying on function derivatives are inapplicable to solve the resultant placement optimization problem. To address this issue, we propose a novel iterative algorithm based on the derivative-free optimization. In each iteration, we first construct a quadratic function to approximate the non-differentiable weighted sum rate under a set of interpolation conditions, and then update the UAVs' placement locations by maximizing the approximate quadratic function subject to a trust region constraint. Numerical results show the convergence of the proposed algorithm. It is also shown that the proposed algorithm achieves a weighted sum rate close to the optimal design based on exhaustive search with much lower implementation complexity, and it significantly outperforms the conventional optimization method based on simplified LoS channel models and the heuristic design with each UAV hovering above its associated GBS. 
\end{abstract}
\section{Introduction}
Unmanned aerial vehicle (UAV)-enabled wireless communications have emerged as a potential key technology for beyond fifth-generation (B5G) and sixth-generation (6G) networks, in which UAVs can either act as aerial users to enable long-range applications (e.g., aerial package delivery), or serve as aerial base stations (BSs) to enhance coverage and increase capacity for ground users \cite{UAVtuto,UAVtuto2,UAV3}. Thanks to the highly controllable mobility and agility, optimizing the UAV placement or trajectory becomes a new design degree of freedom for enhancing the communication performance for both network-connected UAV users and UAV BSs (see, e.g., \cite{ZYInitial,QQ,lipeiming2,QQ2,lixinmin,Valiulahi,lipeiming}). In the literature, there has been a large body of prior works investigating the placement and trajectory optimization in the single-UAV case by considering different setups such as relay channels \cite{ZYInitial}, broadcast channels\cite{QQ}, and multiple access channels \cite{lipeiming2}. Nevertheless, as future B5G/6G networks are envisioned to incorporate a large number of UAVs, the investigation of multi-UAV wireless networks becomes a hot research topic that has attracted growing research interest recently \cite{QQ2,lixinmin,Valiulahi,lipeiming}.

\indent Different from the single-UAV case, the multi-UAV wireless networks face various new technical challenges. In particular, as air-to-ground (A2G) wireless channels usually have strong line-of-sight (LoS) components, the interference among different UAVs may become more severe than that in the conventional terrestrial networks. To deal with this issue, various prior works utilized the techniques of multi-UAV placement \cite{lixinmin,Valiulahi} and trajectory \cite{QQ2} optimization to mitigate the inter-UAV interference, while enhancing the communication link quality. Despite of such progress, these prior works mainly assumed LoS or probabilistic LoS channel models, based on which the placement and trajectory are designed by using conventional convex and non-convex optimization techniques. However, these simplified LoS or statistical A2G channel models generally cannot reflect the reality in complex environment or only describe wireless channels in an average sense. Therefore, these models cannot capture the site- and location-specific channel propagation environments, e.g., due to the blockage and shadowing caused by buildings and vegetation. The over-simplification of channel models may lead to degraded UAV communication performance in practice. 
\\\indent This paper exploits the channel knowledge map (CKM) \cite{CKM} or radio map \cite{S.Zhang,S.Bi, R.Levie, LWJ} to realize environment-aware UAV communications. CKM is a site-specific database that contains location-specific channel-related information for facilitating the wireless system design. Despite of its advantage, in general the channel information (e.g., power gains) from such a CKM cannot be characterized by any analytic functions with respect to the transceivers' locations, thus making the communication system design (e.g., placement and trajectory design for UAVs) very difficult. In the literature, there have been a handful of works studying the UAV placement \cite{mo} and trajectory design \cite{junting.c,Y.Huang,ZY2} in simplified scenarios with only one UAV, by using the techniques of exhaustive search \cite{mo} or reinforcement learning \cite{Y.Huang,ZY2}, or simplifying the general radio map as a segmented channel model \cite{junting.c}. However, these prior designs are inapplicable for our considered scenario with multiple UAVs under the general CKM consideration. This thus motivates us to propose new design techniques for placement optimization in multi-UAV wireless networks.  
\\\indent Specifically, we study the CKM based placement optimization in a multi-UAV wireless network, in which multiple UAV users flying at a fixed altitude send individual messages to their associated ground BSs (GBSs) by using the same frequency band. We assume that the CKM is obtained individually for each BS, which corresponds to a discrete database that contains the site-specific channel gain information for uniformly located sample points. Due to the lack of analytic functions for channel gains, the conventional convex or non-convex optimization methods are not applicable in this case. To tackle this issue, we adopt a novel method called derivative-free optimization \cite{marazzi2002wedge}, which iteratively constructs a series of quadratic functions to approximate the objective function  under a set of interpolation conditions, and accordingly updates the optimization variable by maximizing the approximate function subject to a trust region constraint. By properly designing the trust region, the convergence of the proposed algorithm is ensured. Numerical results show that the proposed algorithm based on derivative-free optimization achieves a performance close to the optimal exhaustive search with a much lower complexity. Furthermore, it is shown that the proposed algorithm significantly outperforms the conventional optimization method based on simplified LoS models and the heuristic design with each UAV hovering above its associated GBS.
\section{System Model}
\begin{figure}[t]
        \centering
        \includegraphics[scale=0.41]{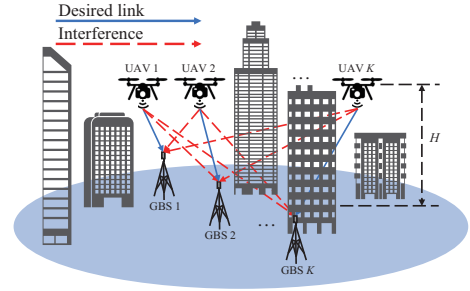}
        \vspace{-11pt}
        \caption{Illustration of the multi-UAV wireless networks.}
        \label{system}
\end{figure}

We consider a multi-UAV wireless network as shown in Fig. \ref{system}, in which $K$ UAV users send individual messages to their respectively associated GBSs over the same frequency band. Let ${\cal K}\triangleq\{1,..., K\}$ denote the set of UAV users or GBSs. Each GBS $k \in {\cal K}$ is located at fixed location $(\hat{x}_k, \hat{y}_k, \hat{H})$ in a three-dimensional (3D) coordinate system, where $\hat{H} \geq 0$ in meters (m) denotes the GBSs' height, and $\mathbf{w}_{k} = (\hat{x}_k, \hat{y}_k)$ denotes the horizontal location. Let $(x_j, y_j, H)$ denote the location of UAV $j\in {\cal K}$, where $\mathbf{q}_{j} = (x_j, y_j)$ denotes the horizontal location of UAV $j$ to be optimized, and $H$ denotes the UAV's altitude. 
\\\indent We assume that one CKM is constructed for each GBS to store the site-specific wireless channel information related to that GBS. In practice, the CKM can be obtained by global channel measurement or constructed approximately by interpolation techniques together with limited measurement data (such as inverse distance weighted, nearest neighbors, and Kriging) \cite{CKM,S.Zhang,S.Bi, R.Levie, LWJ}, and it can also be generated by 3D ray-tracing simulation software (such as Wireless Insite \cite{WI}) based on the 3D environmental map. While the general CKM may contain various channel-related information such as the channel power gain, shadowing, interference, and even angle of arrival/departure (AoA/AoD) \cite{CKM}, in this paper we focus on the specific CKM containing the channel gain information for locations at the constant altitude $H$. Let $\mathcal G_k$ denote the location-specific CKM for GBS $k\in \cal K$, which provides mapping between any given horizontal location $\mathbf{q}$ (at the constant altitude $H$) to the corresponding channel gain $h$, i.e., $\mathbf{q}\in{\mathbb{R}}^{1\times2} \rightarrow h\in{\mathbb{R}}$. Based on the CKM, the channel power gain between each UAV $j \in \cal K $ and GBS $k\in \cal K$ is given by\footnote{The channel power gain here refers to the large-scale channel gain, i.e., path loss and shadowing, since the small-scale fading is hard to be obtained.}
\begin{equation}
    \label{CKM:gain}
    h_{k, j}(\mathbf{q}_{j})=\mathcal G_{k}(\mathbf{q}_{j}),
\end{equation}which is a discrete function without any analytic expression in general. 
Based on the channel power gain in \eqref{CKM:gain}, the received signal-to-interference-plus-noise ratio (SINR) at GBS $k\in{\cal K}$ is given by
\begin{equation}
    \gamma_{k}(\{\mathbf{q}_{j}\})=\frac{P_{k} h_{k, k}(\mathbf{q}_{k})}{{\sum\nolimits_{j \in {\cal K}, j \neq {k}}P_{j} h_{k, j}(\mathbf{q}_{j})} +\sigma^{2}_k},
\end{equation}where $P_{k}$ denotes the transmit power of UAV $k$, and $\sigma^{2}_k$ denotes the noise power at the receiver of GBS $k$, $k\in{\cal K}$. In order to focus our study on the placement optimization, we assume that the transmit power $P_k$'s at different UAVs are given, and leave the optimization of $P_k$'s in future work. By considering Gaussian signalling at each UAV transmitter, the achievable data rate from UAV $k \in{\cal K}$ to its associated GBS $k$ in bits/second/Hertz (bps/Hz) is given by
\begin{equation}
    r_{k}(\{\mathbf{q}_{j}\})=\log _{2}(1+\gamma_{k}(\{\mathbf{q}_{j}\})).
\end{equation}
\indent Our objective is to maximize the weighted sum rate of UAVs. Let $\mv \alpha=\{\alpha_1,...,\alpha_K\}$ denote the predetermined weights that
specify the rate allocation among the $K$ UAVs with $\alpha_k > 0, \forall k \in \cal K$. Then, the weighted sum rate maximization is formulated as
\begin{align}
    \text{(P1)}: &\max\limits_{\{\mathbf{q}_{k}\}}~{\sum\nolimits_{k\in{\cal K}}\alpha_{k}r_{k}(\{\mathbf{q}_{j}\})}\notag\\
    &~~\text {s.t.}~\mathbf{q}_{k} \in \mathcal{A},\forall k \in {\cal K},\label{range}
\end{align}
where $\mathcal{A}$ denotes the target area that limits the locations of the UAVs. Notice that with the environment-aware channel gain offered by CKM, the objective function in problem (P1) is non-differentiable with respect to the UAV locations $\{\mathbf{q}_k\}$. In fact, it even does not have an explicit expression with respect to the UAVs' placement locations. Therefore, problem (P1) is very difficult to be solved, since the conventional convex/non-convex optimization methods are not applicable.
\section{Derivative-Free Placement Optimization for Solving Problem (P1)}
In this section, we propose a novel iterative algorithm to solve problem (P1) based on derivative-free optimization.
\subsection{Derivative-Free Optimization} To start with, we first provide a brief review about derivative-free optimization \cite{marazzi2002wedge}. The purpose of derivative-free optimization is to maximize an objective function $f(\mathbf{x})$ with variable $\mathbf{x} \in \mathbb{R}^{n\times1}$, where the derivative of $f(\mathbf{x})$ is not available. The basic idea is to iteratively approximate $f(\mathbf{x})$ as a series of analytic functions under interpolation conditions, and accordingly update the optimization variable by maximizing the approximate function in each iteration, subject to a trust region constraint. 
\\\indent In particular, consider one iteration $i \ge 1$, in which the local point is denoted by $\mathbf{x}^{(i)}$. First, we construct an analytic function $\phi^{(i)}(\mathbf{x})$ (with $m$ parameters to be determined) to approximate the objective function $f(\mathbf{x})$ near the local point $\mathbf{x}^{(i)}$. In practice, the quadratic and linear functions are widely adopted for constructing $\phi^{(i)}(\mathbf{x})$ \cite{marazzi2002wedge}, which include $m=\frac{1}{2}(n+1)(n+2)$ and $m=n+1$ parameters, respectively. To determine these function parameters, we introduce a so-called \textit{interpolation set} $\mathbf{\Sigma}$ containing $m-1$ points, and accordingly impose the corresponding interpolation conditions, i.e.,
\begin{align}
\phi^{(i)}(\mathbf{y})=f(\mathbf{y}), \forall \mathbf{y} \in \{\mathbf{x}^{(i)}\} \cup \mathbf{\Sigma}.\label{intercond}
\end{align}
Notice that points in the interpolation set are randomly generated initially, and will be updated as the iteration proceeds. Also notice that the constructed interpolation set $\mathbf{\Sigma}$ should be {\it non-degenerate}, i.e., based on the equations in \eqref{intercond} the parameters in the approximate function $\phi^{(i)}(\mathbf{x})$ is non-singular, such that $\phi^{(i)}(\mathbf{x})$ can be uniquely determined.  
\\\indent Next, we update the optimization variable. Towards this end, we first find a trial point as $\mathbf{x}^{(i)}_{+}=\mathbf{x}^{(i)}+\mathbf{s}^{(i)}$, where the update step $\mathbf{s}^{(i)}$ is obtained by maximizing the approximate function $\phi^{(i)}(\mathbf{x}^{(i)}+\mathbf{s})$ subject to a newly imposed trust region, i.e.,
\begin{align}
        \mathbf{s}^{(i)} = \arg&\max\limits_{\mathbf{s}}~\phi^{(i)}(\mathbf{x}^{(i)}+\mathbf{s})\notag\\
       &~~\text {s.t.} ~~~ \|\mathbf{s}\| \leq \Delta,\notag
\end{align}
where $\Delta$ is the trust region size that is set to be a given value $\Delta_0$ at the beginning of the algorithm. If the resultant function value increases (i.e., $f(\mathbf{x}^{(i)}_+) >  f(\mathbf{x}^{(i)})$), then we update the variable as $\mathbf{x}^{(i+1)} =  \mathbf{x}^{(i)}_+$ for the next iteration; otherwise, we have $\mathbf{x}^{(i+1)} =  \mathbf{x}^{(i)}$. In either case, we update the interpolation set $\mathbf{\Sigma}$ by adding $\mathbf{x}^{(i)}_+$ as a new point and removing an old point that is furthest from $\mathbf{x}^{(i)}_+$, and also decrease $\Delta$ to be $\beta \Delta$ with $0< \beta <1$.

Notice that the above iteration terminates when the local point and interpolation set converges (i.e., $\| \mathbf{x}^{(i+1)} - \mathbf{y} \| \le \epsilon, \forall \mathbf{y} \in \mathbf{\Sigma}$) and $\Delta < \epsilon$ is satisfied at the same time, or the maximum number of iterations is met, where $\epsilon$ is a sufficiently small constant threshold for determining the convergence. Nevertheless, if $\Delta < \epsilon $ is met but the local point and interpolation set do not converge yet, then we should reset $\Delta$ as $\Delta = \Delta_0$ and run the next iteration.
\subsection{Derivative-Free UAV Placement Optimization}
\label{UAV:Opti}
Building upon the derivative-free optimization, in this subsection we develop an efficient derivative-free algorithm for solving problem (P1), which is implemented in an iterative manner. For notational convenience, we denote the objective function in problem (P1) as $f(\{\mathbf{q}_{j}\}) = \sum\nolimits_{k\in{\cal K}}\alpha_{k}r_{k}(\{\mathbf{q}_{j}\})$.

In particular, consider any given iteration $i \geq 1$, where the local point is given by $\{\mathbf{q}^{(i)}_j\}$. We denote the locations of the $K$ UAVs $\mathbf{q}^{(i)}_c=(\mathbf{q}^{(i)}_{1},\mathbf{q}^{(i)}_{2},...,\mathbf{q}^{(i)}_{K})^{T} \in \mathbb{R}^{2K\times1}$ for notational convenience.
\\\indent First, we approximate the objective function $f(\{\mathbf{q}_{j}\})$ as approximate function $\phi^{(i)}(\mathbf{q}^{(i)}_{c}+\mathbf{s})$. In particular, we adopt quadratic function for $\phi^{(i)}(\mathbf{q}^{(i)}_{c}+\mathbf{s})$, given by
\begin{equation}
    \phi^{(i)}(\mathbf{q}^{(i)}_{c}+\mathbf{s})=f(\mathbf{q}^{(i)}_{c})+{\mathbf{g}^{(i)}}^T \mathbf{s}+\frac{1}{2} {\mathbf{s}}^{T} \mathbf{G}^{(i)} \mathbf{s},\label{quarfunction}
\end{equation}
where the vector $\mathbf{g}^{(i)} \in \mathbb{R}^{2K\times1}$ and the symmetric matrix $\mathbf{G}^{(i)} \in \mathbb{R}^{2K\times2K}$ contain $m-1=\frac{1}{2}(2K+1)(2K+2)-1$ parameters to be determined. Notice that the consideration of quadratic function in \eqref{quarfunction} is due to the fact that it can properly balance between the approximation performance and the computation burden.
To uniquely determine $\phi^{(i)}(\mathbf{q}^{(i)}_c+\mathbf{s})$, we need to find a non-degenerate interpolation set of $m-1$ points, denoted by
\begin{equation}
    \mathbf{\Sigma}=\{\mathbf{y}_{1}, ..., \mathbf{y}_{m-1}\},
\end{equation}
\text{where} $\mathbf{y}_{l} \in \mathbb{R}^{2K\times1}, ~ l=1, ..., m-1$. Note that points in the interpolation set $\mathbf{\Sigma}$ are obtained through randomly sampling from a uniform distribution over region ${\mathcal A}$ in \eqref{range} and will be updated in each iteration. 
Accordingly, we have the following $m$ equalities:
\begin{equation}
 \label{interpo}
    \phi^{(i)}(\mathbf{q}^{(i)}_{c})=f(\mathbf{q}^{(i)}_{c}), ~ \phi^{(i)}(\mathbf{y}_{l})=f(\mathbf{y}_{l}), ~ l=1, ..., m-1.
\end{equation}
By solving the system of linear equations, we obtain $\mathbf{g}^{(i)}$ and $\mathbf{G}^{(i)}$ and accordingly determine the approximate function $\phi^{(i)}(\mathbf{q}^{(i)}_c+\mathbf{s})$. 

\indent Next, we update the UAVs' placement locations. Towards this end, we first obtain a trial step $\mathbf{s}^{(i)}$, by solving the following problem (P2).
\begin{align}
        \text{(P2)}:&~\max\limits_{\mathbf{s}}~\phi^{(i)}(\mathbf{q}^{(i)}_{c}+\mathbf{s})\notag\\
        &~~~\text {s.t.} ~ \|\mathbf{s}\| \leq \Delta\notag\\
        &~~~~~~~~\mathbf{q}^{(i)}_{c}+\mathbf{s} \in \mathcal{A}.\notag
\end{align}
Note that as the symmetric matrix $\mathbf{G}^{(i)}$ may not be negative semi-definite, problem (P2) is a non-convex quadratic program in general that is difficult to be optimally solved. In order to solve problem (P2), we utilize a trust region subproblem solver tool in \cite{toolbox}, which applies the subroutine Trust Region Step in the BOX (TRSBOX) of BOBYQA algorithm \cite{BOBYQA}. As a result, we approximately obtain the trial step $\mathbf{s}^{(i)}$ and accordingly obtain the trial point $\mathbf{q}^{(i)}_+=\mathbf{q}^{(i)}_{c}+\mathbf{s}^{(i)}$.
\\\indent With the trial step $\mathbf{s}^{(i)}$ at hand, we are ready to update the UAV placement locations $\mathbf{q}_c^{(i+1)}$, together with the trust region size $\Delta$ and the interpolation set $\mathbf{\Sigma}$. Towards this end, we obtain $\mathbf{y}_{l_{\text {out }}}=\arg\max\nolimits_{\mathbf{y} \in \mathbf{\Sigma}}\|\mathbf{y}-\mathbf{q}_{c}^{(i)}\|$ as the point in $\mathbf{\Sigma}$ with the longest distance from the local point $\mathbf{q}^{(i)}_c$, which is a candidate point to be removed from $\mathbf{\Sigma}$. 
In particular, if the trial point $\mathbf{q}^{(i)}_+$ leads to a weighted sum rate $f(\{\mathbf{q}_{k}\})$ that is greater than $\mathbf{q}^{(i)}_c$, i.e., $f(\mathbf{q}^{(i)}_{c}+\mathbf{s}^{(i)})>f(\mathbf{q}^{(i)}_{c})$, then we update the UAV placement locations as $\mathbf{q}^{(i+1)}_c=\mathbf{q}^{(i)}_c+\mathbf{s}^{(i)}$ which is also used as  
the local point in the next iteration $i + 1$; otherwise, we have $\mathbf{q}^{(i+1)}_c=\mathbf{q}^{(i)}_c$.
Furthermore, we update the interpolation set $\mathbf{\Sigma}$ by adding the trial point $\mathbf{q}^{(i)}_+$ as a new point and removing $\mathbf{y}_{l_{\text {out}}}$, and also decrease $\Delta$ by factor $\beta$, i.e., $\Delta \gets \beta \Delta$. 
The above iteration terminates when the local point and interpolation set converges (i.e., $\| \mathbf{q}_c^{(i+1)} - \mathbf{y} \| \le \epsilon, \forall \mathbf{y} \in \mathbf{\Sigma}$) and $\Delta < \epsilon$ is satisfied at the same time, or the maximum number of iterations is met. Nevertheless, if $\Delta < \epsilon$ is met but the local point and interpolation set do not converge yet, then we should reset the trust region size as $\Delta = \Delta_0$ and run the next iteration.
\\\indent In summary, we present the complete algorithm as Algorithm \ref{algorithm}. It is observed that Algorithm 1 results in monotonically non-decreasing objective function values, and the points in the interpolation set $\mathbf{\Sigma}$ will finally converge to the UAV placement locations. Therefore, the convergence of the algorithm can always be ensured. 
\begin{remark}
    It is worth comparing the complexity of the proposed derivative-free placement optimization design versus the optimal exhaustive search benchmark, in which we first sample the interested region ${\mathcal A}$ into $MN$ grids and then compare the weighted sum rates achieved by all the $MN$ possible UAV placement locations to get the desired solution. For the proposed algorithm, the total complexity of constructing the analytic function and solving problem (P2) for updating trial points is $\mathcal{O}(K^{4})$ \cite{BOBYQA}. For exhaustive search, the complexity is $\mathcal{O}(M^KN^{K})$. It is observed that when $K$ becomes large, the complexity of the proposed algorithm is much lower than that of the exhaustive search. 
    \end{remark}
\begin{algorithm}[t]
    \caption{for Solving Problem (P1)}
    \label{algorithm}
    \begin{algorithmic}[1]
    \State Set the initial trust region size $\Delta_{0}>0$, and the initial UAV placement locations $\mathbf{q}^{(0)}_c$. Randomly generate $m-1$ points to compose an initial non-degenerate interpolation set $\mathbf{\Sigma}$. Set the iteration index as $i = 0$ and set the convergence threshold $\epsilon > 0$. 
    \Repeat
    \State Construct the approximate quadratic function $\phi^{(i)}(\mathbf{q}^{(i)}_{c}+\mathbf{s})$ based on interpolation conditions in \eqref{interpo}.
    \State Compute the trial step $\mathbf{s}^{(i)}$ by solving problem (P2), and accordingly obtain the trial point $\mathbf{q}^{(i)}_+=\mathbf{q}^{(i)}_{c}+\mathbf{s}^{(i)}$.
    \State Find $\mathbf{y}_{l_{\text {out }}}=\arg\max\nolimits_{\mathbf{y} \in \mathbf{\Sigma}}\|\mathbf{y}-\mathbf{q}_{c}^{(i)}\|$.
    \If{$f(\mathbf{q}^{(i)}_{c}+\mathbf{s}^{(i)})>f(\mathbf{q}^{(i)}_{c})$}
    \State set $\mathbf{\Sigma}=\{\mathbf{q}^{(i)}_c\} \cup \mathbf{\Sigma} \backslash\{\mathbf{y}_{l_{\text {out }}}\}$, $\mathbf{q}^{(i+1)}_{c}=\mathbf{q}^{(i)}_{c}+\mathbf{s}^{(i)}$.
    \Else
    \State set $\Delta=\beta \Delta$, $\mathbf{q}^{(i+1)}_{c}=\mathbf{q}^{(i)}_{c}$.
    \If{$\|\mathbf{y}_{l_{\text {out }}}-\mathbf{q}^{(i)}_{c}\| \geq \|(\mathbf{q}^{(i)}_{c}+\mathbf{s}^{(i)})-\mathbf{q}^{(i)}_{c}\|$}
    \State set $\mathbf{\Sigma}=\{\mathbf{q}^{(i)}_{c}+\mathbf{s}^{(i)}\} \cup \mathbf{\Sigma} \backslash\{\mathbf{y}_{l_{\text {out }}}\}$.
    \EndIf \State {\textbf{end}}.
    \EndIf \State {\textbf{end}}.
    \If{$\Delta < \epsilon$, and $\| \mathbf{q}^{(i+1)}_c - \mathbf{y} \| >\epsilon, \forall \mathbf{y} \in \mathbf{\Sigma}$}
    \State reset $\Delta = \Delta_0$.
    \EndIf \State {\textbf{end}}.
    \State $i \leftarrow i+1.$
    \Until{$\Delta < \epsilon$, and $\| \mathbf{q}^{(i)}_c - \mathbf{y} \| \leq \epsilon, \forall \mathbf{y} \in \mathbf{\Sigma}$.}
    \end{algorithmic}
    \end{algorithm} 
\section{Numerical Results}
This section presents numerical results to validate the performance of our proposed derivative-free UAV placement optimization design. We consider a specific area in central Shanghai with a size of 300$\times$300 $\text{m}^2$, which consists of a dozen of buildings from a city map database.\footnote{The 3D city map is obtained online from \url{https://www.openstreetmap.org.}} The UAV altitude is set to be $H=50$ m. In the simulation, the received noise power at each GBS $k \in \cal K$ is $\sigma^{2}_k=-100~\mathrm{dBm}$, and the transmit power of each UAV $k \in \cal K$ is $P_k=30~\mathrm{dBm}$. The height of GBSs is set as $\hat{H}=2.0$ m, and the horizontal locations of GBSs are $\mathbf{w}_{1}=(-89.54~\text{m}, 16.30~\text{m})$, $\mathbf{w}_{2}=(-118.22~\text{m}, -53.86~\text{m})$, and $\mathbf{w}_{3}=(-18.15~\text{m}, -80.22~\text{m})$, respectively. The rate weights are set as $\alpha_k=1,\forall k\in{\cal K}$, and thus the sum rate of UAVs is considered as the performance metric. The Remcom Wireless Insite software \cite{WI} is used to generate the CKM dataset at each GBS based on the 3D city map, which provides channel gain values at uniformly distributed points. Each CKM collects channel gains at a total of 4347 points which are uniformly distributed every 5 m in both X- and Y-axis. 
For the proposed algorithm, the decreasing factor of the trust region is adopted as $\beta=0.5$ \cite{marazzi2002wedge}.
\begin{figure}[ht]
    \centering
    \includegraphics[scale=0.4]{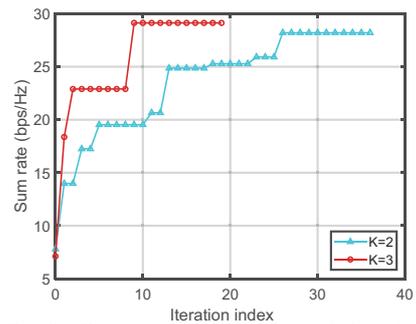}
    \vspace{-14pt}
    \caption{Convergence of the proposed algorithm.}
    \label{convergence}
    \vspace{-7pt}
\end{figure}
\\\indent We consider the exhaustive search and the following two benchmark schemes for performance comparison. For exhaustive search, the candidate UAV locations are uniformly sampled every 5 m, which is consistent with that in the CKM. 
\begin{itemize}
    \item {\it Hovering above GBSs}: Each UAV hovers exactly above its associated GBS with $\mathbf{q}_k = \mathbf{w}_k, \forall k \in \cal K$. This scheme is generally optimal for the special case with $K=1$ or the inter-UAV interference is negligible.
    \item {\it Conventional design with LoS channels}: The LoS path channel model is considered in this scheme, for which the channel power gain between each UAV $j \in \cal K$ and GBS $k \in \cal K$ is given by $ h_{k, j}(\mathbf{q}_{j})={\beta_{0}}/{(\|\mathbf{q}_{j}-\mathbf{w}_{k}\|^{2}+(\hat{H}-H)^2)}$,
    where $\beta_{0}$ denotes the channel gain at a reference distance of $d_0=1$ m and is set as $\beta_0=-30$ dB. This scheme corresponds to solving problem (P1) via conventional non-convex optimization techniques, such as successive convex approximation (see, e.g., \cite{QQ2}).
\end{itemize}
\begin{figure*}[!htb]
    \centering
    \subfigure[CKM of GBS 1.]{
    \includegraphics[scale=0.41]{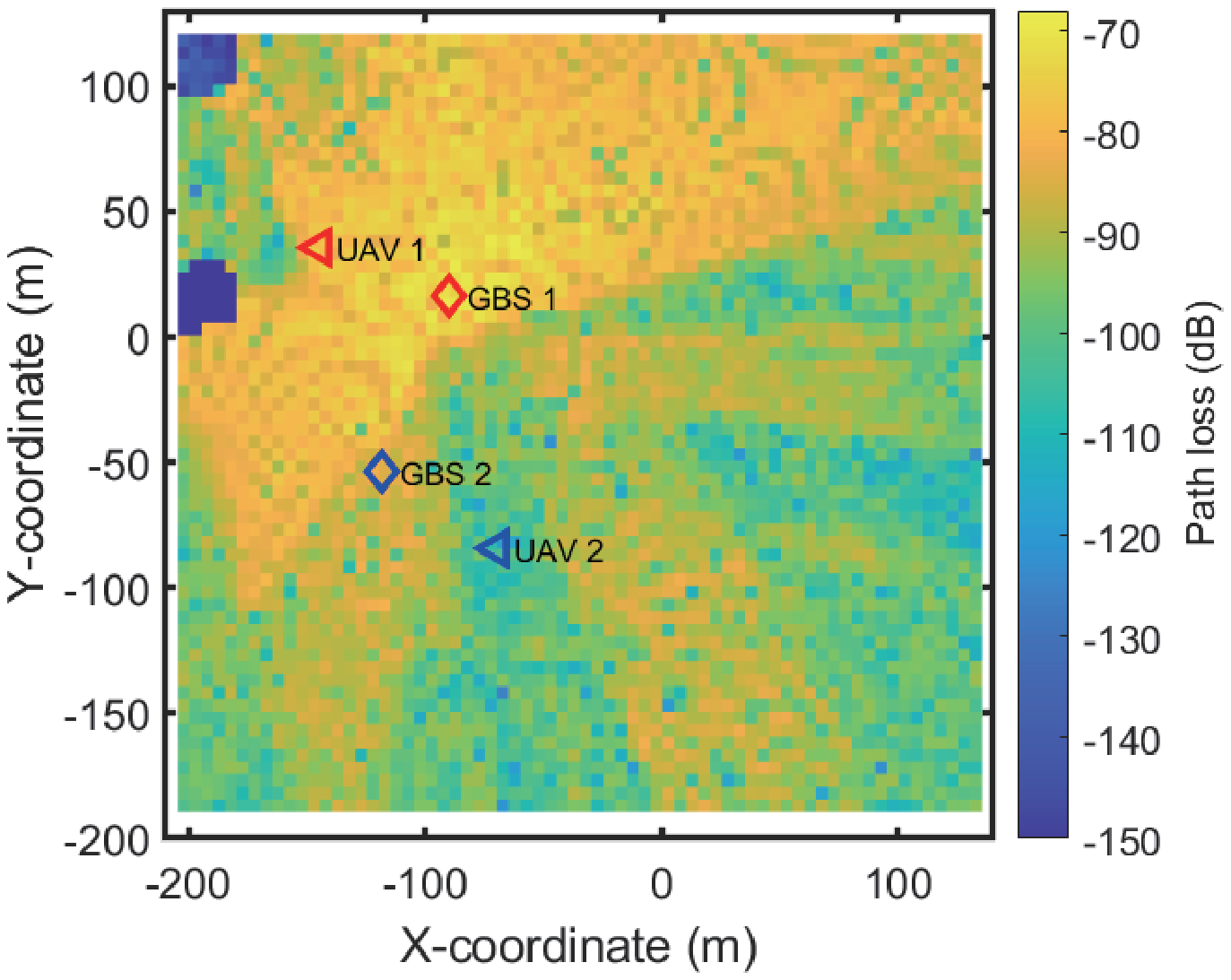}}
    \subfigure[CKM of GBS 2.]{
    \includegraphics[scale=0.41]{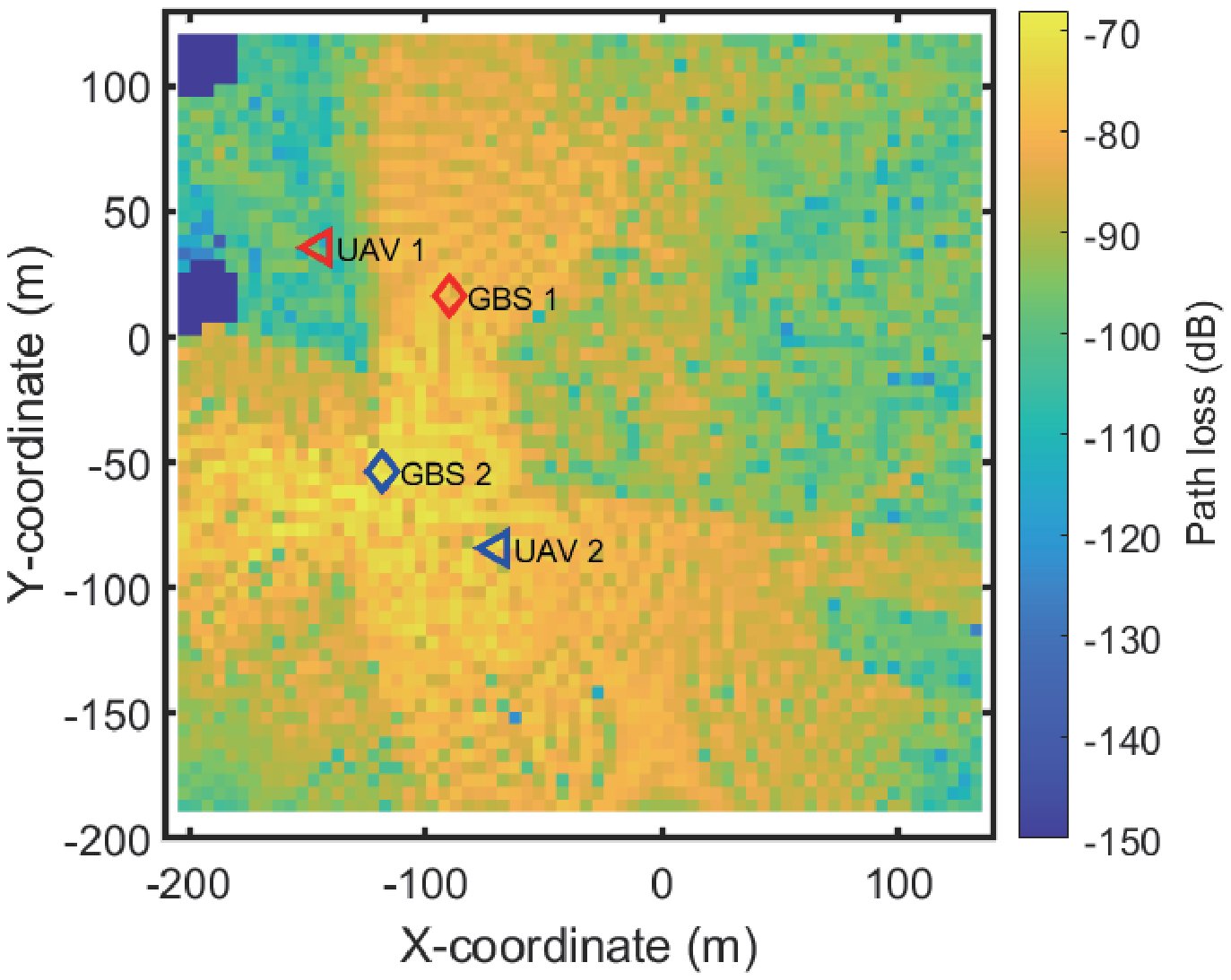}}
    \caption{Optimized UAV horizontal locations for the case with $K=2$, where the diamonds indicate GBS locations, and the triangles indicate UAV locations. }
    \label{fig:CGMs}
\end{figure*}
\begin{figure*}[!htb]
    \centering
    \vspace{-10pt}
    \subfigure[CKM of GBS 1.]{
    \includegraphics[scale=0.41]{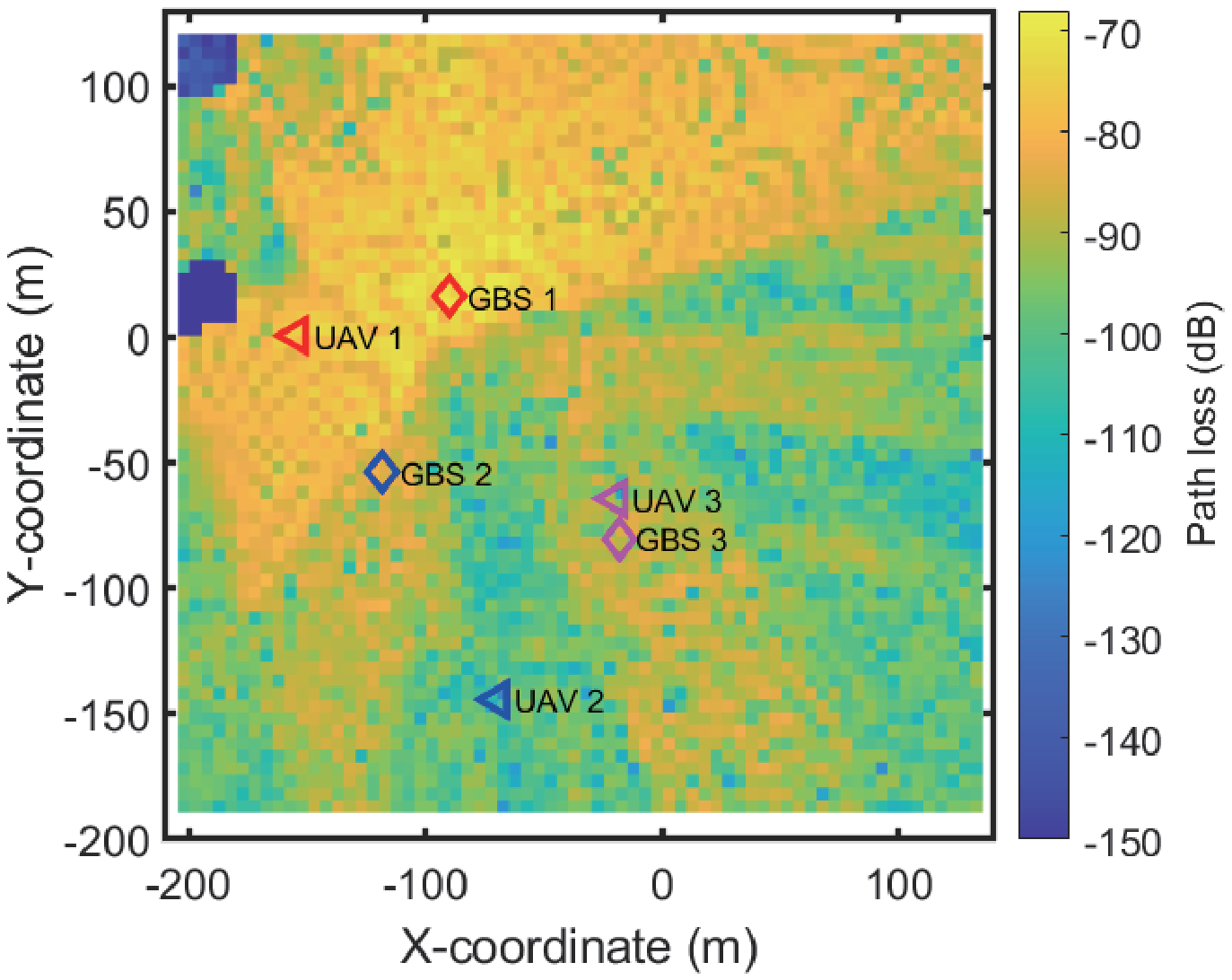}}
    \subfigure[CKM of GBS 2.]{
    \includegraphics[scale=0.41]{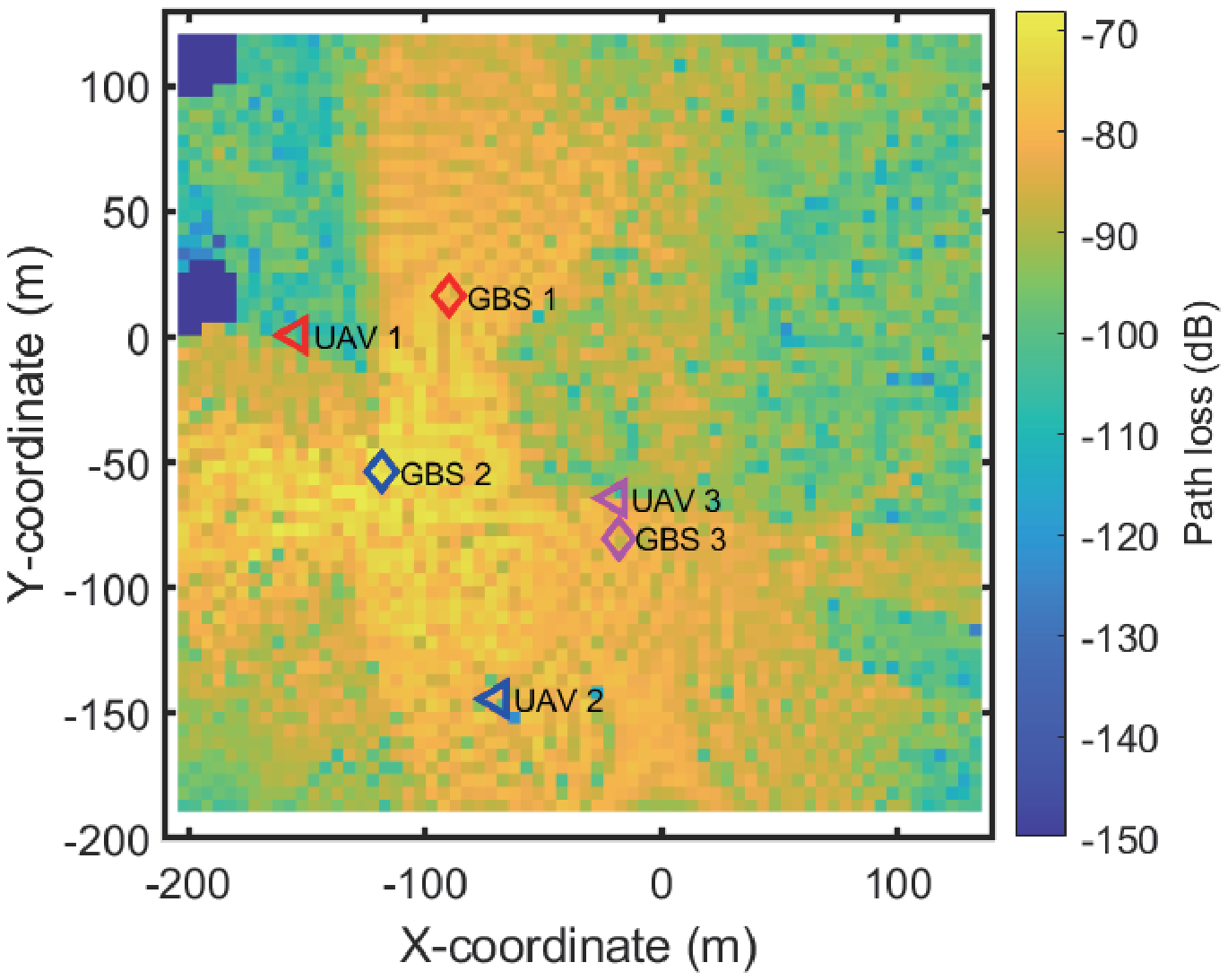}}
    \subfigure[CKM of GBS 3.]{
    \includegraphics[scale=0.41]{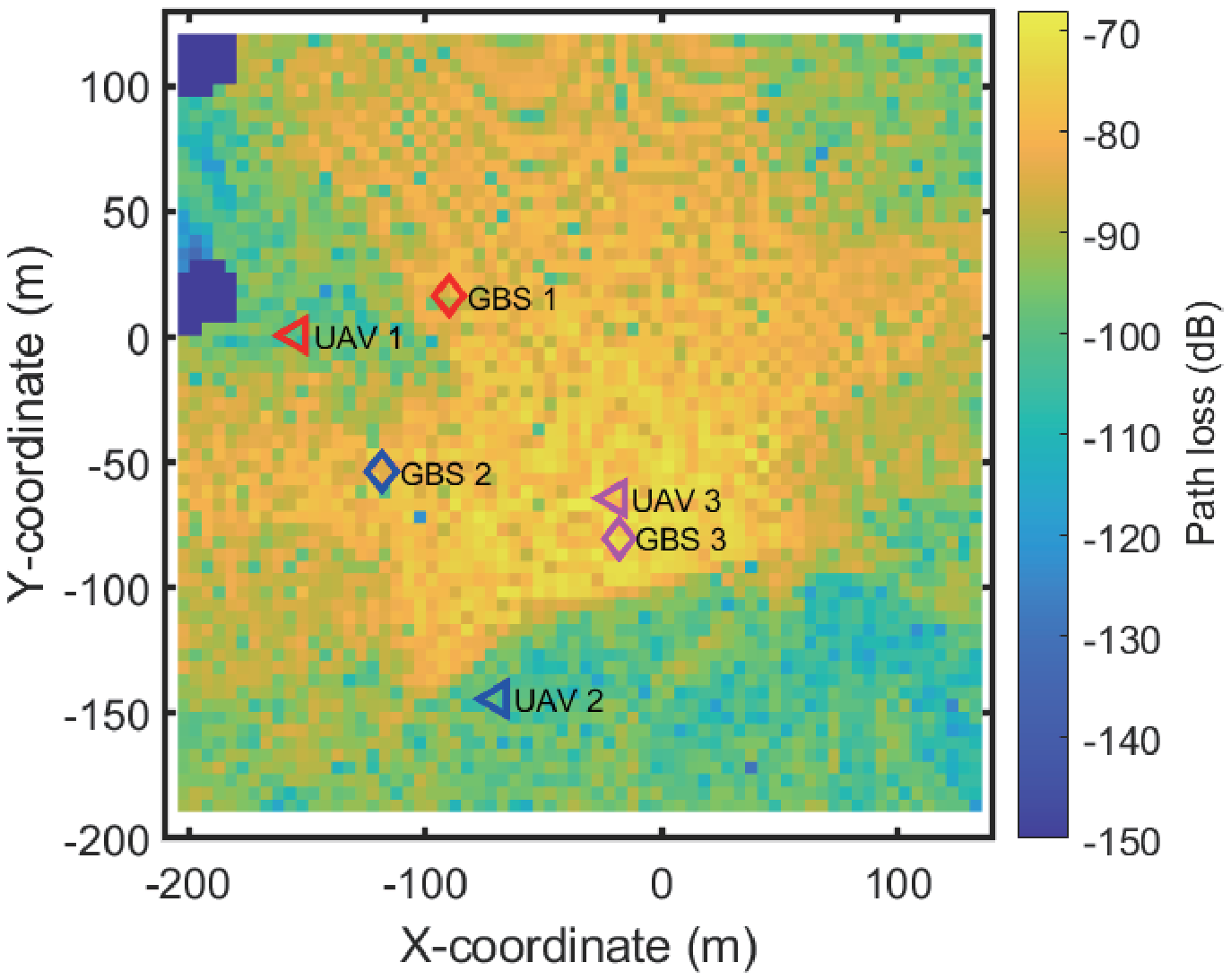}}
    \vspace{-7pt}
    \caption{Optimized UAV horizontal locations for the case with $K=3$, where the diamonds indicate GBS locations, and the triangles indicate UAV locations.}
    \label{fig:CGMs-3}
\end{figure*}
\indent \indent Fig. \ref{convergence} shows the convergence of the proposed algorithm, in both cases with $K = 2$ and $K = 3$. It is observed that the proposed algorithm takes 36 iterations and 19 iterations to converge for the case with $K=2$ and $K=3$, respectively. Furthermore, the proposed algorithm takes about $330$ and $930$ milliseconds to converge for the case with $K=2$ and $K=3$, respectively. For comparison, the {{exhaustive search}} takes about $60$ seconds and 166 hours to find the optimal UAV placement locations for the case with $K=2$ and $K=3$, respectively.
\\ \indent Fig. \ref{fig:CGMs} shows the optimized locations for the case with $K=2$. It is observed that the horizontal location of UAV 1 converges to $\mathbf{q}^{*}_{1}=(-144.64~\text{m}, 35.69~\text{m})$, and that of UAV 2 converges to $\mathbf{q}^{*}_{2}=(-69.64~\text{m}, -84.31~\text{m})$. The sum rate converges to $R=28.2007~\mathrm{bps/Hz}$, with the individual rates $r_1 = 13.5718~\mathrm{bps/Hz}$ and $r_2 = 14.6289~\mathrm{bps/Hz}$ for the two UAV users, respectively. In addition, for the {\it{hovering above GBSs}} scheme, the resulting sum rate is $R=7.7986~\mathrm{bps/Hz}$, with the individual rates $r_1 = 3.4723~\mathrm{bps/Hz}$, and $r_2 = 4.3263~\mathrm{bps/Hz}$. It is observed that if the UAV users hover above their respectively associated GBSs, then both UAVs suffer severe co-channel interference from each other. After placement optimization, each UAV is observed to be placed at a location where its desired link between the corresponding GBS enjoys good channel quality, while its interference link between the other GBS is weak, thus mitigating the co-channel interference to the other UAV thus enhancing the SINR. Furthermore, for {{{exhaustive search}}}, the resulting sum rate is $R=28.2007~\mathrm{bps/Hz}$, which are exactly the same with our proposed algorithm, and the UAV placement locations are also exactly the same.
\\\indent Fig. \ref{fig:CGMs-3} shows the optimized locations for the case with $K=3$ UAVs. It is observed that the horizontal location of UAV 1 converges to $\mathbf{q}^{*}_{1}=(-154.64~\text{m}, 0.69~\text{m})$, and that of UAV 2 and UAV 3 converge to $\mathbf{q}^{*}_{2}=(-69.64~\text{m}, -144.31~\text{m})$ and $\mathbf{q}^{*}_{3}=(-19.64~\text{m}, -64.31~\text{m})$, respectively. The sum rate converges to $R=29.1320~\mathrm{bps/Hz}$, with the individual rates $r_1 = 13.4059~\mathrm{bps/Hz}$, $r_2 = 8.1863~\mathrm{bps/Hz}$, and $r_3 = 7.5398~\mathrm{bps/Hz}$. 
For exhaustive search, the resulting sum rate is $R=30.6819$ bps/Hz, which is very close to the performance achieved by the proposed algorithm. In addition, for the {\it{hovering above GBSs}} scheme, the resulting sum rate is $R=7.1139~\mathrm{bps/Hz}$, with the individual rates $r_1 = 2.7662~\mathrm{bps/Hz}$, $r_2 = 3.1435~\mathrm{bps/Hz}$, and $r_3 = 1.2042~\mathrm{bps/Hz}$.
\begin{figure}[ht]
    \vspace{-7pt}
    \centering
    \includegraphics[scale=0.4]{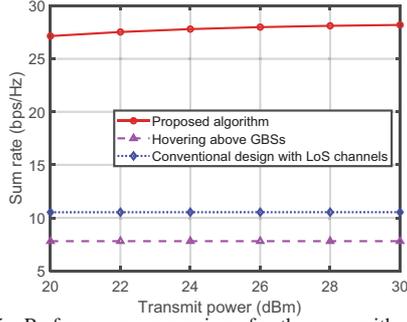}
    \vspace{-14pt}
    \caption{Performance comparison for the case with $K=2$.}
    \label{performance1}
\end{figure}
\begin{figure}[ht]
    \centering
    \includegraphics[scale=0.4]{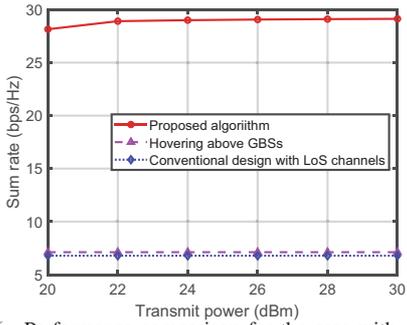}
    \vspace{-14pt}
    \caption{Performance comparison for the case with $K=3$.}
    \label{performance2}
\end{figure}
\\\indent Figs. \ref{performance1} and \ref{performance2} show the sum rate versus the UAV transmit power for the cases with $K=2$ and $K=3$, respectively. It is observed that as the UAV transmit power increases, the sum rate of all UAVs is much higher than that achieved by the {\it{hovering above GBSs}} scheme. Furthermore, compared with our proposed algorithm, the {\it{conventional design with LoS channels}} scheme is observed to achieve poor performance, as it fails to characterize the actual channel environments. This demonstrates that the proposed algorithm with the utilization of CKMs is more accurate in real complex environments.
\section{Conclusion}
In this paper, we considered a multi-UAV wireless network with CKMs, in which we maximized the weighted sum rate by jointly optimizing the UAV placement locations. Note that the CKMs are site-specific discrete databases without any closed-form expression for the channel gains, and thus the considered objective
function is non-differentiable and cannot be solved by conventional convex or non-convex optimization techniques. To tackle this challenge, we proposed a novel iterative algorithm based on derivative-free optimization to obtain a high-quality solution, the key idea of which is to iteratively construct a quadratic function to approximate the objective function. Numerical results showed that our proposed algorithm achieves a close performance to {{exhaustive search}} but with much lower implementation complexity, and also outperforms other benchmark schemes. How to extend the proposed algorithm to other communication scenarios is an interesting and practical issue worth investigating in future work.

\end{document}